\documentclass[3p,times]{elsarticle}
\usepackage{amssymb}

\usepackage{caption}
\usepackage{subfigure}
\usepackage{amsmath}
\usepackage{bm}
\usepackage{color}
\usepackage[colorlinks=True]{hyperref}

\begin{document}
\begin{frontmatter}
\title{A Self-Adaptive Algorithm of the Clean Numerical Simulation (CNS) for Chaos}
\author[label2]{Shijie Qin}
\author[label1,label2,label3]{Shijun Liao \corref{cor1}}
\cortext[cor1]{sjliao@sjtu.edu.cn}

\address[label1]{State Key Laboratory of Ocean Engineering, Shanghai 200240, China}
\address[label2]{Center of Marine Numerical Experiment, School of Naval Architecture, Ocean and Civil Engineering, Shanghai Jiao Tong University, Shanghai 200240, China}
\address[label3]{School of Physics and Astronomy, Shanghai Jiao Tong University, Shanghai 200240, China}

\begin{abstract}
The background numerical noise  $\varepsilon_{0} $ is determined by the maximum of truncation error and round-off error.   For a chaotic system, the numerical error $\varepsilon(t)$ grows exponentially, say, $\varepsilon(t) = \varepsilon_{0} \exp(\kappa\,t)$, where $\kappa>0$ is the so-called noise-growing exponent. This is the reason why one can not gain a convergent simulation of chaotic systems in a long enough interval of time by means of traditional algorithms in double precision, since the background numerical noise $\varepsilon_{0}$ might stop decreasing because of the use of double precision. This restriction can be overcome by means of the clean numerical simulation (CNS), which can decrease the background numerical noise $\varepsilon_{0}$ to any required tiny level. A lot of successful applications show the novelty and validity of the CNS. In this paper, we further propose some strategies to greatly increase the computational efficiency of the CNS algorithms for chaotic dynamical systems. It is highly suggested to keep a balance between truncation error and round-off error and besides to progressively enlarge the background numerical noise $\varepsilon_{0}$, since the exponentially increasing numerical noise $\varepsilon(t)$ is much larger than it. Some examples are given to illustrate the validity of our strategies for the CNS.
\end{abstract}	


\end{frontmatter}

\section{Introduction}

For a chaotic dynamical system, the sensitivity dependence on initial conditions (SDIC) was first discovered by Poincar\'{e} \cite{poincare1890probleme}, and then this phenomenon was discovered once again by Lorenz \cite{lorenz1963deterministic} with a more familiar name ``butterfly-effect''. Due to the SDIC, a very weak, small-scale disturbance of the initial condition will give rise to a huge deviation of numerical solution of the chaotic system after a long enough temporal interval \cite{sprott2003chaos,wang2013forward,kuznetsov2018finite}. Furthermore, it was found that a chaotic dynamical system not only has the sensitivity dependence on initial conditions (SDIC) but also possesses the sensitivity dependence on numerical algorithms (SDNA), as reported by Lorenz \cite{lorenz1989computational, lorenz2006computational}. All of these phenomena are due to the exponential increase of noise (or uncertainty) of chaotic systems, but unfortunately {\em artificial} numerical noises (i.e. truncation errors and round-off errors) are always {\em inevitable} for almost all of the numerical algorithms. Thus, for a chaotic dynamical system, calculated trajectories of computer-generated simulations obtained by means of different numerical algorithms (with single/double precision) and different time steps are mostly quite different. Naturally, such kind of non-replicability/unreliability of chaotic solution has brought plenty of heated debates on the credence of the numerical simulation of chaotic dynamical system \cite{nazare2020note}, and someone even made an extremely pessimistic conclusion that ``for chaotic systems, numerical convergence cannot be guaranteed {\em forever}'' \cite{Teixeira2007Time}. In addition, it has been recently reported that ``shadowing solutions can be almost surely nonphysical'', which ``invalidates the argument that small perturbations in a chaotic system can only have a small impact on its statistical behavior'' \cite{chandramoorthy2021probability}.

To gain a {\em reproducible/reliable} numerical simulation of chaotic systems, Liao \cite{Liao2009} proposed a brand-new numerical strategy, namely the ``Clean Numerical Simulation'' (CNS) \cite{Liao2013, Liao2014, LIAO2014On}, to control the background numerical noise, say, truncation error and round-off error, during a temporal interval $t\in[0,T_c]$, where $T_c$ is the so-called ``critical predictable time'' and this temporal interval should be long enough for calculating statistics. In the frame of the CNS \cite{Liao2009, Liao2013, Liao2014, LIAO2014On, hu2020risks, qin2020influence, xu2021accurate, AAMM-14-799}, the temporal truncation error and the spatial truncation error are able to be decreased to a {\em required} small level via using the Taylor expansion method with a high {\em enough} order in the temporal dimension and adopting a fine {\em enough} discretization method in the spatial dimension (such as the high-order spatial Fourier expansion), respectively. Significantly, all of the physical and numerical variables/parameters should be represented by means of the multiple precision (MP) \cite{oyanarte1990mp} with a large {\em enough} number of significant digits, and thus the round-off error is also able to be decreased to a {\em required} small level. Moreover, an additional numerical simulation of the identical chaotic system with the even smaller numerical noise is required and performed in order to determine such a ``critical predictable time'' $T_c$, so that the numerical noise could be negligible and thus the computer-generated solution of a chaotic system is reproducible/reliable within the whole spatial computational domain and in the temporal interval $[0,T_c]$. In this way, different from some other general numerical algorithms, the CNS is able to give the reproducible/reliable numerical simulation of a chaotic dynamical system within a finite but long enough temporal interval.

Here it should be emphasized that although our CNS strategy is based on the classical Taylor series method \cite{Corliss1982114} as well as the multiple precision \cite{oyanarte1990mp}, the scientific significance of this strategy is mainly about the ``critical predictable time'' $T_c$: the CNS can greatly reduce the background numerical noise, i.e. truncation error and round-off error, to any a required tiny level so that the numerical noise is negligible compared with the ``true'' physical solution, and thus the corresponding numerical result of a chaotic system is reproducible/reliable in an interval of time $[0,T_c]$ that is long enough for statistics, as described in the next section. In other words, the results of chaotic dynamical systems given by the CNS can be regarded as a ``clean'' benchmark solution \cite{hu2020risks, qin2020influence, xu2021accurate, AAMM-14-799}, which is the main purpose of proposing this CNS strategy.
By contrast, solely adopting the Taylor series method \cite{Corliss1982114, chang1994atomft, barrio2005performance, nedialkov2008solving} to solve a chaotic system for high precision, one usually does not focus on the ``critical predictable time'' $T_c$ and thus obtain a mixture of the ``true'' physical solution and the ``false'' numerical noise, which are mostly at the same order of magnitude, since the background numerical noise of a simulation of chaos should increase exponentially (and quickly) until to the same level of ``true'' physical solution, which is not considered by traditional numerical strategies.

For the computer-generated simulation of a chaotic dynamical system given by a certain numerical algorithm, it is well-known that the averaged level of numerical noise should increase exponentially within a temporal interval $[0,T_c]$, i.e.
\begin{equation}
\varepsilon(t)=\varepsilon_0 \exp(\kappa\,t), \hspace{1.0cm} t\in[0,T_c],    \label{noise_exp}
\end{equation}
where the noise-growing exponent $\kappa$ is a positive constant (usually corresponding to the maximum Lyapunov exponent for a chaotic dynamical system with a finite degree of freedom), $T_c$ is the above-mentioned critical predictable time in the frame of the CNS strategy, $\varepsilon_0$ represents the level of initial/background numerical noise (which is determined by the initial truncation error and the initial round-off error), and $\varepsilon(t)$ denotes the averaged level of evolving noise for a computer-generated simulation, respectively. Considering that there might be another increasing pattern of numerical noise \cite{2018Linear, 2020Superfast} that is more meticulous, the exponential growing (\ref{noise_exp}) is still suitable for a long-time simulation. Theoretically, the critical predictable time $T_c$ is determined by a given value of the critical numerical noise $\varepsilon_c$, i.e. $\varepsilon_c=\varepsilon_0 \exp(\kappa\,T_c)$ that leads to
\begin{equation}
T_c = \frac{1}{\kappa}\ln\left(\frac{\varepsilon_c}{\varepsilon_0}\right).    \label{noise_exp_c}
\end{equation}
Obviously, if the value of $\varepsilon_c$ is unchanged, the smaller the level of the initial/background numerical noise $\varepsilon_0$, the larger
the critical predictable time $T_c$.

Unfortunately, it is impossible in practice to obtain the evolving noise $\varepsilon(t)$ with high accuracy, because we do not know the true (physical) solution of a numerical simulation of chaos. Thus, a practical approach with satisfied numerical precision is required to calculate the $\varepsilon(t)$. Let $\mathbf{x}\in\Omega$ represent the dimensional vector in a chaotic dynamical system, $\phi(\mathbf{x},t)$ denote the solution of numerical (computer-generated) simulation that is reproducible/convergent within $t\in[0,T_c]$ possessing the initial/background numerical noise $\varepsilon_0$, and $\phi'(\mathbf{x},t)$ denote another solution (using the identical initial/boundary conditions and physical parameters) that is reliable within $t\in[0,T_c']$ possessing the initial/background numerical noise $\varepsilon_0'$ which is smaller than $\varepsilon_0$. Due to the exponentially growing property (\ref{noise_exp}) of numerical noise for a chaotic dynamical system, there is $T_c'>T_c$ and that $\phi'(\mathbf{x},t)$ within $t\in[0,T_c]$ must be superior and much closer to the physical solution (true solution) compared with $\phi(\mathbf{x},t)$. And thus, $\phi'(\mathbf{x},t)$ could be seen as a benchmark solution to help us determine the numerical noise of $\phi(\mathbf{x},t)$ within $\mathbf{x}\in\Omega$ approximately. Therefore, practically, the evolving noise $\varepsilon(t)$ is obtained via comparing $\phi(\mathbf{x},t)$ (possessing the background numerical noise $\varepsilon_0$) with a superior numerical solution $\phi'(\mathbf{x},t)$ (possessing the smaller background numerical noise $\varepsilon_0'$).

Up to now, the above-mentioned CNS strategy has been applied to many chaotic dynamical systems successfully with the corresponding computer-generated simulations being reproducible and of course reliable. For example, via using some general numerical algorithms with {\em double} precision, one always obtains the reproducible numerical solutions of the well-known Lorenz system only in a short temporal interval, i.e. $t\in[0,32]$ approximately \cite{Liao2009}. By contrast, via using the CNS strategy, a reproducible/convergent numerical simulation of the same chaotic Lorenz system within quite a long temporal interval, i.e. $t\in[0, 10000]$, was obtained {\em for the first time} by Liao and Wang \cite{LIAO2014On}. Besides, Liao and Li \cite{liao2015inherent} studied the evolution of the microscopic physical uncertainty of initial condition for the famous three-body system (which is chaotic) by means of the CNS, and they found that the uncertainty finally becomes macroscopical, which leads to the random escape of the three-body system as well as the behavior of symmetry breaking. It indicates that the uncertainty of microscopic physics could be the origin of large-scale randomness for the well-known three-body system. Furthermore, the numerical noise of the CNS strategy can be controlled to be much smaller than the uncertainty of microscopic physics, and thus via using the CNS, Lin et al. \cite{lin2017origin} theoretically provided rigorous evidence to demonstrate that the microscopic thermal fluctuation should be the origin of large-scale randomness of the two-dimensional turbulent Rayleigh-B{\'e}nard convection.
Significantly, with the help of China's national supercomputer, the CNS strategy was applied to investigate the periodic orbits of the famous three-body problem, and more than $2000$ brand-new families of periodic orbits were discovered successfully by Li et al. \cite{Li2017More, li2018over, li2019collisionless}. Those newly found periodic orbits were reported twice in the popular magazine {\em New Scientist} \cite{NewScientist2017, NewScientist2018}, because, for the three-body problem, there are only three families of chaotic periodic orbits that had ever been found since Newton mentioned this famous problem three hundred years ago!
It is also worth noting that, according to a known periodic orbit as well as three equal masses, and integrating the governing equations by means of the CNS, Li et al. \cite{li2021one} obtained $135445$ brand-new periodic orbits of arbitrarily unequal masses of the three-body system, including $13315$ stable ones.
In addition, using the CNS in quite a long temporal interval, Xu et al. \cite{xu2021accurate} obtained the reliable/reproducible trajectories of a free-fall disk that is chaotic under some certain physical parameters, and the CNS strategy is able to help him accurately forecast the position and posture of the chaotic free-fall disk near the bifurcation point.

As for spatiotemporal chaos, Hu \& Liao \cite{hu2020risks} and Qin \& Liao \cite{qin2020influence} proposed an efficient CNS strategy utilized in physical space to numerically solve the 1D complex Ginzburg-Landau equation (CGLE) and the damped driven sine-Gordon equation (SGE), respectively, which further demonstrates the effectiveness of the CNS strategy that can exactly maintain both the statistical properties and symmetric features of the spatiotemporal chaotic systems in which general numerical algorithms with double precision always fail.
Recently, taking the CNS strategy as a tool, it has been found that the statistical features (such as the probability density function) of some chaotic dynamical systems are extremely sensitive to the tiny noise/disturbance, and thus this kind of chaos is called ultra-chaos by Liao \& Qin \cite{AAMM-14-799}. As a brand-new concept, the ultra-chaos might deepen and enrich our understandings about chaos and turbulence.
Furthermore, with the help of CNS, Qin \& Liao \cite{qin_liao_2022} provide rigorous evidence that numerical noises as a kind of tiny artificial stochastic disturbances have quantitatively and qualitatively large-scale influences on a sustained turbulence.
In a word, the above-mentioned investigations demonstrate the effectiveness and potential of the CNS for complex chaotic dynamical systems.

Although the CNS is able to be applied to obtain the reproducible/convergent numerical simulation of a chaotic dynamical system within a long enough temporal interval, it is more time-consuming compared with some other general numerical algorithms with double precision \cite{LIAO2014On, lin2017origin}. In this paper, according to the exponentially growing property of noise $\varepsilon(t)$ in (\ref{noise_exp}), we propose a modified strategy of the CNS, called the ``self-adaptive CNS'', to significantly increase the computational {\em efficiency} of the CNS algorithm. To illustrate its validity, we apply the CNS with the self-adaptive precision to some chaotic systems, such as the Lorenz equation, the hyper-chaotic R\"{o}ssler system, the three-body problem, and the damped driven sine-Gordon equation.

\section{Basic ideas of the self-adaptive CNS}    \label{sec_Lorenz}

In this section, let us use the Lorenz equations \cite{lorenz1963deterministic}
\begin{equation}
\left\{
\begin{array}{l}
\dot{x}(t)=\sigma\,[y(t)-x(t)],    \\
\dot{y}(t)=R\,x(t)-y(t)-x(t)\,z(t),  \\
\dot{z}(t)=x(t)\,y(t)+b\,z(t),   \\
\end{array}
\right.  \label{lorenz_eq}
\end{equation}
in the case of
\begin{equation}
\sigma=10, \;\; R=28, \;\;  b=-8/3,    \label{lorenz_parameters}
\end{equation}
under the initial condition
\begin{equation}
x(0)=-15.8, \;\; y(0)=-17.48, \;\;  z(0)=35.64,    \label{lorenz_ini}
\end{equation}
as one of the most famous chaotic systems (with one positive Lyapunov exponent) to briefly describe the basic ideas of the self-adaptive CNS.

The CNS algorithm for the Lorenz system (\ref{lorenz_eq})-(\ref{lorenz_ini}) is mainly based on the $M$th-order Taylor series in the temporal interval $[t, t+\Delta t]$:
\begin{equation}
x(t+\Delta t)\approx x(t)+\sum_{m=1}^{M}x^{[m]}(t)\,(\Delta t)^m,    \label{Taylor_lorenz_x}
\end{equation}
\begin{equation}
y(t+\Delta t)\approx y(t)+\sum_{m=1}^{M}y^{[m]}(t)\,(\Delta t)^m,    \label{Taylor_lorenz_y}
\end{equation}
\begin{equation}
z(t+\Delta t)\approx z(t)+\sum_{m=1}^{M}z^{[m]}(t)\,(\Delta t)^m,    \label{Taylor_lorenz_z}
\end{equation}
where $\Delta t$ is the time step and
\begin{equation}
x^{[m]}(t)=\frac{1}{m!}\frac{d^mx(t)}{dt^m}, \;\; y^{[m]}(t)=\frac{1}{m!}\frac{d^my(t)}{dt^m}, \;\;  z^{[m]}(t)=\frac{1}{m!}\frac{d^mz(t)}{dt^m}
\end{equation}
are the high-order temporal derivatives.
Differentiating both sides of Eqs.~(\ref{lorenz_eq}) $(m-1)$ times with respect to $t$ and then dividing them by $m!$, we obtain the iterative formulae
\begin{equation}
x^{[m]}(t)=\frac{\sigma}{m}\,\left[y^{[m-1]}(t)-x^{[m-1]}(t)\right],    \label{Taylor_lorenz_Dx}
\end{equation}
\begin{equation}
y^{[m]}(t)=\frac{1}{m}\,\left[R\,x^{[m-1]}(t)-y^{[m-1]}(t)-\sum_{i=0}^{m-1}x^{[i]}(t)\,z^{[m-1-i]}(t)\right],    \label{Taylor_lorenz_Dy}
\end{equation}
\begin{equation}
z^{[m]}(t)=\frac{1}{m}\,\left[\sum_{i=0}^{m-1}x^{[i]}(t)\,y^{[m-1-i]}(t)+b\,z^{[m-1]}(t)\right],    \label{Taylor_lorenz_Dz}
\end{equation}
for arbitrary $m\geq1$. Note that parallel technology can be applied to calculate the sum terms in (\ref{Taylor_lorenz_x})-(\ref{Taylor_lorenz_z}), (\ref{Taylor_lorenz_Dy}) and (\ref{Taylor_lorenz_Dz}).

According to (\ref{noise_exp_c}), the background numerical noises $\varepsilon_{0}$ must be small enough if one needs a reliable (reproducible) chaotic solution within a large temporal interval $t\in[0,T_{c}]$. It is worth noting that the background numerical noises $\varepsilon_{0}$ in (\ref{noise_exp_c}) is a constant, which is determined by the maximum of the spatio-temporal truncation error (resulting from the truncation of an infinite number of series) and the round-off error (resulting from a limited number of significant digits of data). The basic idea of the above-mentioned CNS algorithm is to greatly decrease both the temporal truncation error and round-off error so that the background numerical noise $\varepsilon_{0}$ is small enough for a numerical simulation to be reproducible/reliable within a given temporal interval $t\in[0,T_{c}]$.

Obviously, if the temporal Taylor expansion (\ref{Taylor_lorenz_x})-(\ref{Taylor_lorenz_z}) is given a large {\em enough} order $M$, the temporal truncation error is able to be decreased under a {\em required} small level. More importantly, different from other traditional algorithms, we express all of the physical and numerical variables/parameters by means of the multiple precision (MP) via choosing the significant digits with a large {\em enough} number $N_s$, and thus the round-off error is also able to be decreased under a {\em required} small level. In this way, both the temporal truncation error and round-off error are able to be decreased under a {\em required} small level via the CNS.

Note that computer-generated solutions of chaotic Lorenz system (\ref{lorenz_eq})-(\ref{lorenz_ini}) given by some general numerical algorithms with {\em double} precision are reproducible/convergent in quite a short temporal interval $t\in[0,32]$. In $2014$, Liao \& Wang \cite{LIAO2014On} obtained a reproducible/convergent numerical simulation $(x,y,z)$ of the above-mentioned Lorenz system in quite a long temporal interval $t\in[0,10000]$ (Lorenz unit time) by means of a parallel algorithm of the CNS using the $3500\hspace{0.2mm}$th-order Taylor expansion ($M=3500$) with the constant time step $\Delta t=0.01$ and $4180$-digit multiple precision ($N_s=4180$) for all physical and numerical variables/parameters, whose reproducibility/reliability (from the mathematical viewpoint) was confirmed by means of another simulation $(x',y',z')$ given by the CNS with the smaller background numerical noise using the $3600\hspace{0.2mm}$th-order Taylor expansion ($M=3600$) with the time step $\Delta t = 0.01$ and $4515$-digit multiple precision ($N_s=4515$). For simplicity, define the relative error
\begin{equation}
\varepsilon(t)=\frac{\big|x'(t)-x(t)\big|+\big|y'(t)-y(t)\big|+\big|z'(t)-z(t)\big|}{\big|x'(t)\big|+\big|y'(t)\big|+\big|z'(t)\big|},    \label{lorenz_relative_error}
\end{equation}
where $(x,y,z)$ and $(x',y',z')$ are the two CNS results mentioned above. It was found that (\ref{noise_exp}) indeed holds with the noise-growing exponent $\kappa\approx0.91$ (which corresponds to the maximum Lyapunov exponent of this Lorenz system), say, $\kappa/\ln\hspace{-0.3mm}10\approx0.40$, indicating that the background numerical noise $\varepsilon_{0}$ will be enlarged nearly $10^{4000}$ times at $t=10000$. This is the reason why Liao \& Wang \cite{LIAO2014On} had to use the $4180$-digit multiple precision ($N_s=4180$) and the $3500\hspace{0.2mm}$th-order Taylor expansion ($M=3500$) in their CNS algorithm so as to greatly decrease the background numerical noise $\varepsilon_{0}$ to a very tiny level!

Frankly speaking, one hardly uses 3500th-order Taylor expansion and data in a multiple precision with 4180 significant digits in practice. However, from a theoretical viewpoint, the {\em reproducible/convergent} chaotic simulation of the Lorenz equations in such a long interval of time is very important, since it gives, {\em for the first time}, direct evidence that one can indeed gain a {\em reproducible/convergent} trajectory of chaotic systems in a long enough interval of time. It invalidates the argument that ``for chaotic systems, numerical convergence cannot be guaranteed {\em forever}'' \cite{Teixeira2007Time}, although a large number of calculations are required: it took 220.9 hours (i.e. about 9 days and 5 hours) using 1200 CPUs of the National Supercomputer TH-1A at Tianjian, China \cite{LIAO2014On}. This kind of convergent simulation in $t\in[0,10000]$ can be used as a benchmark solution of the chaotic Lorenz system (\ref{lorenz_eq})-(\ref{lorenz_ini}) to verify the modified CNS algorithms, as described below.

How to increase the computational efficiency of the CNS?

\subsection{Keeping a balance between truncation error and round-off error}

Note that the background numerical noises $\varepsilon_{0}$ in (\ref{noise_exp}) is determined by the maximum of the truncation error and round-off error.
So, for solving this problem, it is the optimum that the temporal truncation error is at the same level as the round-off error. So, we should keep a balance between the temporal truncation error and the round-off error. Unlike Liao and Wang \cite{LIAO2014On} who used a constant time step,
the {\em variable} stepsize (VS) strategy \cite{Barrio2005} is able to be applied to the above-mentioned CNS algorithm by means of an allowed tolerance $tol$ (whose value is given) of the governing equations. Referring to Barrio {\em et al.} \cite{Barrio2005}, the optimal time stepsize is given by
\begin{equation}
\Delta t=min\hspace{0.2mm}\left(\frac{tol^{\frac{1}{M}}}{\|x_i^{[M-1]}(t)\|_\infty^{\frac{1}{M-1}}},
\frac{tol^{\frac{1}{M+1}}}{\|x_i^{[M]}(t)\|_\infty^{\frac{1}{M}}}\right),    \label{tol}
\end{equation}
where $M$ denotes the order of Taylor expansion, $tol$ denotes the allowed tolerance, $\|~\|_\infty$ is the infinite norm for the variable $x_i$ ($i=1,2,3$), and $x_1(t)$, $x_2(t)$, $x_3(t)$ correspond to $x(t)$, $y(t)$, $z(t)$, respectively. Considering that the parallel technology is applied to calculate the sum terms in (\ref{Taylor_lorenz_x})-(\ref{Taylor_lorenz_z}), (\ref{Taylor_lorenz_Dy}) and (\ref{Taylor_lorenz_Dz}), here we use the empirical formula
\begin{equation}
M=\left\lceil-1.5\log_{10}(tol)\right\rceil,    \label{lorenz_M}
\end{equation}
to determine a proper order of Taylor expansion for the high calculating efficiency. Furthermore, considering the round-off error of data should be controlled at the same level of the temporal truncation error, we choose
\begin{equation}
tol=10^{-N_s},    \label{lorenz_tol}
\end{equation}
where $N_{s}$ denotes the number of significant digits chosen by means of the multiple precision. In this way, we can control the background numerical noise by means of choosing the number $N_s$ for multiple precision and keeping a balance between the temporal truncation error and round-off error via (\ref{lorenz_M}) and (\ref{lorenz_tol}) with an optimal value of the time step via (\ref{tol}).

The above-mentioned strategy can greatly increase the computational efficiency of the CNS algorithm. For the chaotic Lorenz system (\ref{lorenz_eq})-(\ref{lorenz_ini}), according to (\ref{noise_exp}), here $\varepsilon_{0} = 10^{-N_{s}}$ and $\kappa\approx0.91$ (i.e. $\kappa/\ln\hspace{-0.3mm}10\approx0.40$), we should choose $N_{s} = 4020$ so as to guarantee that the numerical noise $\varepsilon(t)$ is nearly at the level of $10^{-20}$ at $t = 10000$. In fact, using $N_{s} = 4020$ and the corresponding $tol = 10^{-N_{s}}=10^{-4020}$ and $M=\left\lceil-1.5\log_{10}(tol)\right\rceil=\left\lceil1.5\hspace{0.3mm}N_{s}\right\rceil=6030$, we obtain a convergent simulation by means of a parallel CNS algorithm using 1200 CPUs of the National Supercomputer TH-2 at Guangzhou, China, which agrees in the accuracy of more than 20 significant digits in the whole interval of time $t\in[0,10000]$ with the benchmark solution given by Liao \& Wang \cite{LIAO2014On}. Note that it took only 96.8 hours (i.e. about 4 days and 1 hours), just about 44\% of the CPU time required by Liao \& Wang \cite{LIAO2014On} in a supercomputer. This illustrates that the computational efficiency of the CNS algorithm can be indeed greatly increased by using an optimal time step (\ref{tol}) and keeping a balance between the truncation error and round-off error, as mentioned above.

\subsection{Using self-adaptive multiple-precision}

The background numerical noise $\varepsilon_{0}$ is determined by the maximum of the truncation error and round-off error. According to (\ref{noise_exp}), one had to use very small background numerical noise $\varepsilon_{0}$ so as to gain a convergent chaotic simulation in a long interval of time. This is indeed true. For example, to gain the convergent benchmark solution of the chaotic Lorenz system in $t\in[0,10000]$, Liao \& Wang \cite{LIAO2014On} used the $3500\hspace{0.2mm}$th-order Taylor expansion ($M=3500$) with the time step $\Delta t = 0.01$ in the $4180$-digit multiple precision ($N_s=4180$). The corresponding background numerical noise is indeed rather small. But, unfortunately, it is rather time-consuming.

Note that a key point of the CNS is to determine the critical predictable time $T_c$. Since there exists a balance between the truncation error and round-off error, for the chaotic Lorenz system (\ref{lorenz_eq})-(\ref{lorenz_ini}) in the last section, it is reasonable to assume that the background numerical noise should be equal to the round-off error, say, $\varepsilon_{0} = 10^{-N_{s}}$, where $N_s$ is the initial significant digit number of the multiple-precision (MP). Then, according to (\ref{noise_exp}) we have
\begin{equation}
\varepsilon(t) = \varepsilon_{0} \exp(\kappa\,t) = 10^{-(N_{s}-\kappa\,t/\ln\hspace{-0.3mm}10)}, \hspace{1.0cm} t\in[0,T_c],    \label{noise_Ns_exp}
\end{equation}
where the noise-growing exponent $\kappa\approx0.91$ is known and is generally equal to the leading Lyapunov exponent of a temporal chaos, and further
\begin{equation}
\varepsilon_c = 10^{-N_{s}+\kappa\,T_c/\ln\hspace{-0.3mm}10},
\end{equation}
which gives the relationship between the initial significant digit number $N_s$ of the multiple-precision (MP) and the critical predictable time $T_c$:
\begin{equation}
N_s = \left\lceil \frac{\gamma\,\kappa\,T_c}{\ln\hspace{-0.3mm}10}-\log_{10}\varepsilon_c \right\rceil,    \label{Ns-Tc}
\end{equation}
where $\varepsilon_c$ denotes the critical numerical noise that is close to the order of magnitude of the true physical solution, $\lceil\cdot\rceil$ stands for the ceiling function, $\gamma\geq1$ is a constant used here as a kind of safety factor, respectively.

According to (\ref{noise_Ns_exp}), the numerical noise $\varepsilon(t)$ increases {\em exponentially}. Thus, after a short time such as at $t=t'$, $\varepsilon(t')$ becomes much larger than the background numerical noise $\varepsilon_{0}$. So, it is {\em unnecessary} to keep the background numerical noise $\varepsilon_{0}$ being the {\em same} in the whole interval $t\in[0,T_{c}]$. In theory, according to (\ref{noise_exp}), using a larger background numerical noise $\varepsilon'_{0}$ does not influence the numerical result in $t\geq t'$, as long as $\varepsilon_{0}<\varepsilon'_{0}\leq\varepsilon(t)$. Note that, for the above-mentioned CNS algorithm of the Lorenz system (\ref{lorenz_eq})-(\ref{lorenz_ini}), the {\em larger} background numerical noise corresponds to the multiple-precision with a {\em smaller} number $N_{s}$ of significant digits and a {\em larger} allowed tolerance $tol = 10^{-N_{s}}$ that further leads to a {\em smaller} order $M=\left\lceil-1.5\log_{10}(tol)\right\rceil=\left\lceil1.5\hspace{0.3mm}N_{s}\right\rceil$ of Taylor expansion.

Thus, according to (\ref{Ns-Tc}), after integrating a temporal interval $t\in[0,t')$, where $t'<T_c$, it is sufficient to use a smaller number of significant digits
\begin{equation}
N_s = \left\lceil \frac{\gamma\,\kappa\,(T_c-t')}{\ln\hspace{-0.3mm}10}-\log_{10}\varepsilon_c \right\rceil,    \label{SA-Ns}
\end{equation}
and a {\em larger} allowed tolerance $tol = 10^{-N_{s}}$ that further leads to a {\em smaller} order $M=\left\lceil-1.5\log_{10}(tol)\right\rceil=\left\lceil1.5\hspace{0.3mm}N_{s}\right\rceil$ of Taylor expansion, respectively, to gain the CNS result for $t\geq t'$. In practice, it is unnecessary to change $N_s$, $tol$ and the corresponding $M$ at each time step but at some given times such as $t'=100$, $500$, $1000$ and so on.

\begin{table}[t]
\renewcommand\arraystretch{1.5}
\tabcolsep 0pt
\caption{CPU times of the self-adaptive CNS algorithm with the adjustable multiple-precision (MP) and an optimal variable time step for the chaotic Lorenz system (\ref{lorenz_eq})-(\ref{lorenz_ini}) in $t\in[0,10000]$, i.e. $T_c=10000$, where the number $N_{s}$ of significant digits is determined by (\ref{SA-Ns-prac}) with the allowed tolerance $tol = 10^{-N_{s}}$ that further leads to the order $M=\left\lceil-1.5\log_{10}(tol)\right\rceil=\left\lceil1.5\hspace{0.3mm}N_{s}\right\rceil$ of Taylor expansion with the optimal time step (\ref{tol}), taking the safety factor $\gamma=1.1$ and using different values of $\Delta T$.}
\vspace*{-7pt}\label{lorenz_CPU}
\begin{center}
\def\temptablewidth{0.5\textwidth}
{\rule{\temptablewidth}{1pt}}
\begin{tabular*}{\temptablewidth}{@{\extracolsep{\fill}}cc}
~~$\Delta T$ & CPU time (hours)~~  \\
\hline
~~25 & 37.7~~  \\
~~50 & 37.2~~  \\
~~100 & 37.4~~  \\
~~500 & 40.3~~  \\
~~1000 & 42.6~~  \\
\end{tabular*}
{\rule{\temptablewidth}{1pt}}
\end{center}
\end{table}

Substituting $\kappa=0.91$ into (\ref{SA-Ns}) and choosing $\varepsilon_c=10^{-2}$, we have the following relationship
\begin{equation}
N_s = \left\lceil \frac{\gamma\,(T_c-t^*)}{2.53}+2 \right\rceil\approx\left\lceil \frac{\gamma\,(T_c-t^*)}{2.53} \right\rceil,    \label{SA-Ns-prac}
\end{equation}
where $t^*=n\,\Delta T$ with $n=0,1,2,...$ and $\Delta T$ being a constant such as $\Delta T=25$, $50$, $100$, $500$, $1000$ and so on. In practice, there is $T_c-t^*>500$ for the high enough remaining precision, say, the value of $N_s$ is stopped decreasing when $t>9500$ for the long time simulation with $t\in[0,10000]$ in this section. Taking the safety factor $\gamma=1.1$ and using different values of $\Delta T$, the corresponding CPU times of this self-adaptive CNS algorithm with the adjustable multiple-precision (MP) and an optimal variable time step, for the chaotic Lorenz system (\ref{lorenz_eq})-(\ref{lorenz_ini}) in $t\in[0,10000]$, i.e. $T_c=10000$, are listed in Table~\ref{lorenz_CPU}. It indicates that the required CPU time of the above-mentioned self-adaptive CNS algorithm is not very sensitive to the value of $\Delta T$, and thus we can choose $\Delta T=0.5\%\hspace{0.3mm}T_c=50$ for the relatively higher computational efficiency.

\begin{table}[t]
\renewcommand\arraystretch{1.5}
\tabcolsep 0pt
\caption{Convergent result of the chaotic Lorenz system (\ref{lorenz_eq})-(\ref{lorenz_ini}) in $t\in[0,10000]$, i.e. $T_c=10000$, given by the self-adaptive CNS algorithm using $tol = 10^{-N_{s}}$ and $M=\left\lceil-1.5\log_{10}(tol)\right\rceil=\left\lceil1.5\hspace{0.3mm}N_{s}\right\rceil$ with the optimal time step (\ref{tol}), where $N_{s}$ is determined by (\ref{SA-Ns-prac}) with taking $\gamma=1.1$ and using $\Delta T=100$.}
\vspace*{-7pt}\label{lorenz_results}
\begin{center}
\def\temptablewidth{1.0\textwidth}
{\rule{\temptablewidth}{1pt}}
\begin{tabular*}{\temptablewidth}{@{\extracolsep{\fill}}cccc}
$t$ & $x$ & $y$ & $z$ \\
\hline
1000 & 13.881997000862393623 & 19.918303160406394373 & 26.901943308376105536 \\
2000 & -6.8738836932050180481 & -1.4848348276698421977 & 31.349521074674276721 \\
3000 & 1.6932902170011335241 & 3.6003418650451083164 & 21.410875101298497293 \\
4000 & -7.6926663606916323997 & -13.499590676622338604 & 14.199428882538458225 \\
5000 & -6.0844510954990075032 & -10.813737089458431017 & 12.739116756422288312 \\
6000 & 0.21673563458354078642 & 2.1042785739999677006 & 22.124608735478140521 \\
7000 & -11.394859731998057561 & -16.575389386215504779 & 23.681268415272744261 \\
8000 & -1.2658734776208739301 & -2.3362702560379947755 & 17.495968339114928401 \\
9000 & 13.479653230046728502 & 17.282101858684218362 & 29.238196888213967777 \\
10000 & -15.817277998368267071 & -17.366868329556944701 & 35.558386165882592794 \\
\end{tabular*}
{\rule{\temptablewidth}{1pt}}
\end{center}
\end{table}

The CNS algorithm described in \S~2.1, together with the above-mentioned self-adaptive strategy, can greatly increase the computational efficiency of the CNS. For example, for the chaotic Lorenz system (\ref{lorenz_eq})-(\ref{lorenz_ini}) in $t\in[0,10000]$, using $tol = 10^{-N_{s}}$ and $M=\left\lceil-1.5\log_{10}(tol)\right\rceil=\left\lceil1.5\hspace{0.3mm}N_{s}\right\rceil$ with the optimal time step (\ref{tol}), where $N_{s}$ is determined by (\ref{SA-Ns-prac}) with taking $\gamma=1.1$, $T_c=10000$ and $\Delta T=50$, we successfully obtain a reproducible/convergent numerical simulation by means of the parallel CNS algorithm using 1200 CPUs of the National Supercomputer TH-2 at Guangzhou, China, which agrees with the benchmark solution given by Liao \& Wang \cite{LIAO2014On} in the accuracy of at least 20 significant digits in the whole interval of time $t\in[0,10000]$, as shown in Table~\ref{lorenz_results}.
Note that it took only 37.2 hours (i.e. about 1 day and 13 hours), just about 17\% of the CPU time of the previous CNS algorithm applied by Liao \& Wang \cite{LIAO2014On} who likewise used a supercomputer. Thus, the self-adaptive CNS algorithm mentioned above has indeed much higher computational efficiency than the previous CNS with the constant background numerical noise.

\begin{figure}[h!]
    \begin{center}
        \begin{tabular}{cc}
            \includegraphics[width=3.2in]{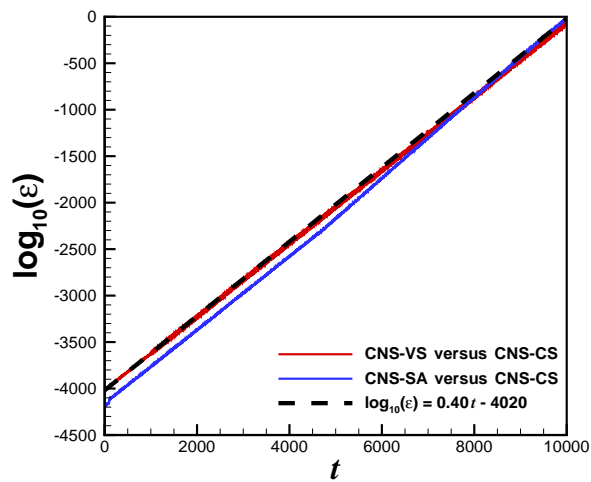}
        \end{tabular}
    \caption{Evolving noises $\varepsilon(t)$ of the CNS results of the Lorenz system (\ref{lorenz_eq})-(\ref{lorenz_ini}) in the whole interval of time $t\in[0,10000]$, given by the CNS algorithm combined with the variable stepsize strategy (marked by CNS-VS, red solid line) using $N_s=4020$, $tol=10^{-N_s}=10^{-4020}$, $M=\left\lceil-1.5\log_{10}(tol)\right\rceil=\left\lceil1.5\hspace{0.3mm}N_{s}\right\rceil=6030$ with the optimal time step (\ref{tol}), and given by the self-adaptive CNS algorithm (marked by CNS-SA, blue solid line) using $tol = 10^{-N_{s}}$, $M=\left\lceil-1.5\log_{10}(tol)\right\rceil=\left\lceil1.5\hspace{0.3mm}N_{s}\right\rceil$ with the optimal time step (\ref{tol}), where $N_{s}$ is determined by (\ref{SA-Ns-prac}) with taking $\gamma=1.1$, $T_c=10000$ and $\Delta T=100$. These evolving noises $\varepsilon(t)$ are obtained via the comparison with the previous CNS algorithm applied by Liao \& Wang \cite{LIAO2014On} that has the constant stepsize (marked by CNS-CS). Black dashed line: $\log_{10}(\varepsilon)=0.40\hspace{0.2mm}t-4020$.}    \label{lorenz_noise_T10000}
    \end{center}
\end{figure}

In addition, the numerical precision of the above-mentioned self-adaptive CNS algorithm can be guaranteed: as shown in Fig.~\ref{lorenz_noise_T10000}, the results given by this strategy (marked by CNS-SA) and the CNS algorithm combined with the variable stepsize strategy described in \S~2.1 (marked by CNS-VS) are in the accuracy of at least 20 significant digits in the whole interval of time $t\in[0,10000]$, which is obtained via the comparison with the previous CNS algorithm applied by Liao \& Wang \cite{LIAO2014On} that has the constant stepsize and the constant background numerical noise (marked by CNS-CS).

\section{Some examples}

\subsection{Self-adaptive CNS for hyper-chaotic R\"{o}ssler system}    \label{sec_Rossler}

The chaotic four-dimensional R\"{o}ssler system \cite{rossler1979equation}
\begin{equation}
\left\{
\begin{array}{l}
\dot{x}(t)=-\,y(t)-z(t),    \nonumber \\
\dot{y}(t)=x(t)+a\,y(t)+w(t),    \nonumber \\
\dot{z}(t)=b+x(t)\,z(t),      \nonumber \\
\dot{w}(t)=-\,c\,z(t)+d\,w(t),     \nonumber \\
\end{array}
\right.  \label{rossler_x}
\end{equation}
in the case of
\begin{equation}
a=0.25, \;\; b=3, \;\; c=0.5, \;\; d=0.05,    \label{rossler_parameters}
\end{equation}
under  the initial condition
\begin{equation}
x(0)=-20, \;\; y(0)=z(0)=0, \;\;  w(0)=15,    \label{rossler_ini}
\end{equation}
has aroused wide concern as a typical hyper-chaotic system \cite{rossler1979equation, zhang2005controlling, stankevich2020scenarios}, since it has two positive Lyapunov exponents. How can we gain a convergent chaotic simulation of R\"{o}ssler system (\ref{rossler_x})-(\ref{rossler_ini}) in the accuracy of 20 significant digits in $t\in[0,10000]$?

The CNS algorithm for the hyper-chaotic R\"{o}ssler system (\ref{rossler_x})-(\ref{rossler_ini}) is also based on the $M$th-order truncated Taylor series in the temporal interval $[t, t+\Delta t]$:
\begin{equation}
x(t+\Delta t)\approx x(t)+\sum_{m=1}^{M}x^{[m]}(t)\,(\Delta t)^m,    \label{Taylor_rossler_x}
\end{equation}
\begin{equation}
y(t+\Delta t)\approx y(t)+\sum_{m=1}^{M}y^{[m]}(t)\,(\Delta t)^m,    \label{Taylor_rossler_y}
\end{equation}
\begin{equation}
z(t+\Delta t)\approx z(t)+\sum_{m=1}^{M}z^{[m]}(t)\,(\Delta t)^m,    \label{Taylor_rossler_z}
\end{equation}
\begin{equation}
w(t+\Delta t)\approx w(t)+\sum_{m=1}^{M}w^{[m]}(t)\,(\Delta t)^m,    \label{Taylor_rossler_w}
\end{equation}
where the high-order derivatives are governed by
\begin{equation}
x^{[m]}(t)=\frac{1}{m}\,\left[-\,y^{[m-1]}(t)-z^{[m-1]}(t)\right],    \label{Taylor_rossler_Dx}
\end{equation}
\begin{equation}
y^{[m]}(t)=\frac{1}{m}\,\left[x^{[m-1]}(t)+a\,y^{[m-1]}(t)+w^{[m-1]}(t)\right],    \label{Taylor_rossler_Dy}
\end{equation}
\begin{equation}
z^{[m]}(t)=\frac{1}{m}\,\left[B_m+\sum_{i=0}^{m-1}x^{[i]}(t)\,z^{[m-1-i]}(t)\right],    \label{Taylor_rossler_Dz}
\end{equation}
\begin{equation}
w^{[m]}(t)=\frac{1}{m}\,\left[-\,c\,z^{[m-1]}(t)+d\,w^{[m-1]}(t)\right],    \label{Taylor_rossler_Dw}
\end{equation}
for arbitrary $m\geq1$ and
\begin{equation}
B_m=
\left\{
\begin{array}{lr}
b, \hspace{1.0cm} m=1,    \\
0, \hspace{1.0cm} m>1.
\end{array}
\right.
\end{equation}
Note that the parallel technology can be applied to calculate the sum terms in (\ref{Taylor_rossler_x})-(\ref{Taylor_rossler_w}) and (\ref{Taylor_rossler_Dz}).

\begin{table}[t]
\renewcommand\arraystretch{1.5}
\tabcolsep 0pt
\caption{Convergent result of the hyper-chaotic R\"{o}ssler system (\ref{rossler_x})-(\ref{rossler_ini}) in $t\in[0,10000]$ given by the CNS parallel algorithm.}
\vspace*{-7pt}\label{rossler_results}
\begin{center}
\def\temptablewidth{1.0\textwidth}
{\rule{\temptablewidth}{1pt}}
\begin{tabular*}{\temptablewidth}{@{\extracolsep{\fill}}ccccc}
$t$ & $x$ & $y$ & $z$ & $w$ \\
\hline
1000  &  -33.992602 &  -5.5093173 &  0.087878252  &  20.503330 \\
2000  &  -13.578396 &  9.9097524 &  0.23536319  &  13.972979 \\
3000  &  -76.551626 &  32.540905 &  0.039413405  &  40.392583 \\
4000  &  -27.968165 &  -19.878204 &  0.10477112  &  24.712276 \\
5000  &  -21.983158 &  20.457146 &  0.14308917  &  23.787920 \\
6000  &  -11.968879 &  20.979201 &  0.32526220  &  26.481453 \\
7000  &   -5.9175355 &  11.379490 &  1.0511723  &  17.005655 \\
8000  &  -18.606119 &  -8.2632834 &  0.15776751  &  17.709076 \\
9000  &  -16.668563 &  13.935457 &  0.18978718  &  30.452723 \\
10000  &  -56.166749 &  27.803911 &  0.053903377  &  29.797559 \\
\end{tabular*}
{\rule{\temptablewidth}{1pt}}
\end{center}
\end{table}

It is easy for us to know that the maximum Lyapunov exponent of the hyper-chaotic R\"{o}ssler system (\ref{rossler_x})-(\ref{rossler_ini}) is about 0.11, which gives us the corresponding noise-growing exponent $\kappa \approx 0.11$ in (\ref{noise_exp}). In this case, if $\varepsilon_{0} = 10^{-N_{s}}$, we have the numerical noise evolution
\[ \varepsilon(t) \approx \varepsilon_{0} \exp(0.11\hspace{0.3mm}t) \approx 10^{-(N_{s}-0.048\hspace{0.3mm}t)}. \]
If our CNS simulation should be in the accuracy of at least 8 significant digits in the whole interval of $t\in[0,10000]$, we have $T_{c} = 10000$ and
\[  -(N_{s} - 0.048 \hspace{0.3mm} T_{c}) \leq -8,  \]
which gives $N_{s} \geq 488$, indicating that we should choose $N_{s} = 488$.

Similarly, in the frame of the CNS, the background numerical noise (i.e. truncation error and round-off error) of this system can be decreased under a {\em required} tiny level by means of choosing a large {\em enough} order $M$ of the Taylor expansion (\ref{Taylor_rossler_x})-(\ref{Taylor_rossler_w}) and a large {\em enough} number $N_s$ of significant digits for multiple-precision. First, following Liao \& Wang \cite{LIAO2014On} who used a constant time step, we obtain a reproducible/convergent simulation of the hyper-chaotic R\"{o}ssler system (\ref{rossler_x})-(\ref{rossler_ini}) in $t\in[0,10000]$ by means of a parallel CNS algorithm using the 415th-order Taylor expansion ($M=415$) with a fixed time step $\Delta t = 0.01$ in the multiple precision of 488 significant digits ($N_{s} = 488$), as listed in Table~\ref{rossler_results}. In fact, this convergent simulation result agrees in the accuracy of more than 8 significant digits in the whole interval of time $t\in[0,10000]$ compared with the benchmark solution given by another CNS using the 500th-order Taylor expansion ($M=500$) with a fixed time step $\Delta t = 0.01$ in the multiple precision of 550 significant digits ($N_{s} = 550$). It takes 5804 seconds (i.e. about 1 hours and 37 minutes) using 50 Intel's CPUs (Xeon Silver 4114) on our local cluster.

Then, we apply the strategy of keeping a balance between truncation error and round-off error (mentioned in \S~2.1) to increase the computational efficiency of the CNS algorithm. The variable stepsize (VS) scheme is applied with an optimal time step determined by (\ref{tol}), where $i=1,2,3,4$ is for this hyper-chaotic R\"{o}ssler system and thus $x_1(t)$, $x_2(t)$, $x_3(t)$, $x_4(t)$ correspond to $x(t)$, $y(t)$, $z(t)$, $w(t)$, respectively. Considering that the parallel technology is applied to calculate the sum terms in (\ref{Taylor_rossler_x})-(\ref{Taylor_rossler_w}) and (\ref{Taylor_rossler_Dz}), here we adopt the empirical formula
\begin{equation}
M=\left\lceil-1.5\log_{10}(tol)\right\rceil,    \label{rossler_M}
\end{equation}
to choose a proper order of Taylor expansion for the high calculating efficiency. Besides, (\ref{lorenz_tol}) is used to keep the balance between the round-off error and the truncation error. In this way, we can control the background numerical noise $\varepsilon_{0}$ by choosing the number $N_s$ of significant digits for multiple-precision, say, $\varepsilon_{0}$ is at the level of $10^{-N_{s}}$.

By means of the CNS algorithm described in \S~2.1 using a {\em fixed} value of $N_{s} = 488$ for multiple-precision, $tol = 10^{-488}$ for the allowed tolerance, $M=\left\lceil-1.5\log_{10}(tol)\right\rceil=\left\lceil1.5\hspace{0.3mm}N_{s}\right\rceil=732$ for the order of Taylor expansion, and an optimal time step given by (\ref{tol}), we obtain a reproducible/convergent numerical simulation of the hyper-chaotic R\"{o}ssler system (\ref{rossler_x})-(\ref{rossler_ini}) in $t\in[0,10000]$, which gives exactly the same result as those listed in Table~\ref{rossler_results}. And it takes 608 seconds (i.e. about 10 minutes) using 50 Intel's CPUs (Xeon Silver 4114) on our local cluster, say, only 10.5\% CPU time of the previous CNS algorithm (i.e. 5804 seconds) with a {\em fixed} time step. The convergence of this CNS result is confirmed by comparing it with another CNS result with the even smaller background numerical noise, given by a fixed value of $N_{s} = 550$ for multiple-precision, $tol = 10^{-550}$ for the allowed tolerance, $M=\left\lceil-1.5\log_{10}(tol)\right\rceil=\left\lceil1.5\hspace{0.3mm}N_{s}\right\rceil=825$ for the order of Taylor expansion, and the optimal time step via (\ref{tol}).

\begin{figure}[h!]
    \begin{center}
        \begin{tabular}{cc}
            \includegraphics[width=3.2in]{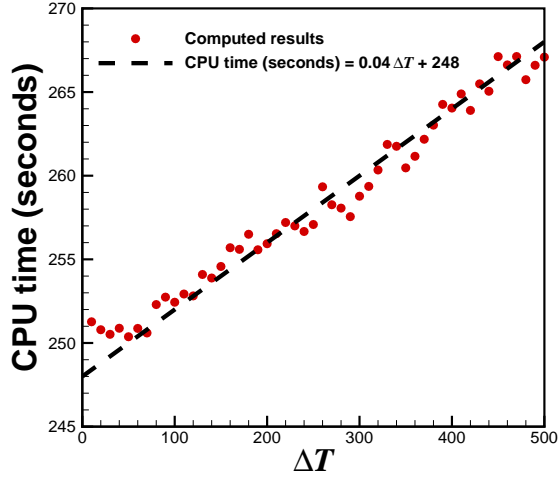}
        \end{tabular}
    \caption{CPU times of the self-adaptive CNS algorithm for the hyper-chaotic R\"{o}ssler system (\ref{rossler_x})-(\ref{rossler_ini}) in $t\in[0,10000]$, given by different values of $\Delta T$, using $tol = 10^{-N_{s}}$ and $M=\left\lceil-1.5\log_{10}(tol)\right\rceil=\left\lceil1.5\hspace{0.3mm}N_{s}\right\rceil$ with the optimal time step (\ref{tol}), where $N_{s}$ is determined by (\ref{SA-Ns-prac-EX1}) with taking $\gamma=1.1$ and $T_c=10000$. Red circle: computed results; Black dashed line: CPU time (seconds) equals to $0.04\hspace{0.3mm}\Delta T+248$.}    \label{CPU_tao_rossler}
    \end{center}
\end{figure}

In addition, the computational efficiency can be further increased by means of the self-adaptive CNS algorithm described in \S~2.2. Substituting $\kappa=0.11$ into (\ref{SA-Ns}) and choosing $\varepsilon_c=10^{-2}$, we have the following relationship
\begin{equation}
N_s = \left\lceil \frac{\gamma\,(T_c-t^*)}{20.9}+2 \right\rceil\approx\left\lceil \frac{\gamma\,(T_c-t^*)}{20.9} \right\rceil,    \label{SA-Ns-prac-EX1}
\end{equation}
where $t^*=n\,\Delta T$ with the non-negative integer $n$. In practice, there is $T_c-t^*>1000$ for the high enough remaining precision, say, the value of $N_s$ is stopped decreasing when $t>9000$ for the long time simulation with $t\in[0,10000]$ in this section. Using different values of $\Delta T$, the corresponding CPU times of this self-adaptive CNS algorithm with the adjustable multiple-precision (MP) and an optimal variable time step, for the hyper-chaotic R\"{o}ssler system (\ref{rossler_x})-(\ref{rossler_ini}) in $t\in[0,10000]$, i.e. $T_c=10000$, are illustrated in Fig.~\ref{CPU_tao_rossler}.
It indicates that there is an approximate linear relationship that the required CPU time (seconds) equals to $0.04\hspace{0.3mm}\Delta T+248$. Since the corresponding slope 0.04 is rather small, it reconfirms the conclusion that the required CPU time of the above-mentioned self-adaptive CNS algorithm is not very sensitive to the value of $\Delta T$. Thus, in practice we can choose $\Delta T=0.5\%\hspace{0.3mm}T_c=50$ for the relatively higher computational efficiency.

For the hyper-chaotic R\"{o}ssler system (\ref{rossler_x})-(\ref{rossler_ini}) in $t\in[0,10000]$, using $tol = 10^{-N_{s}}$ and $M=\left\lceil-1.5\log_{10}(tol)\right\rceil=\left\lceil1.5\hspace{0.3mm}N_{s}\right\rceil$ (according to (\ref{lorenz_tol}) and (\ref{rossler_M}), respectively) with the optimal time step (\ref{tol}), where $N_{s}$ is determined by (\ref{SA-Ns-prac-EX1}) with taking $\gamma=1.1$, $T_c=10000$ and $\Delta T=50$, we successfully obtain the {\em same} reproducible/convergent numerical simulation by means of the parallel CNS algorithm together with the above-mentioned self-adaptive strategy using 50 Intel's CPUs (Xeon Silver 4114) on our local cluster, as listed in Table~\ref{rossler_results}, which agrees with the benchmark solution given by another CNS, using the 500th-order Taylor expansion ($M=500$) with a fixed time step $\Delta t = 0.01$ in the multiple precision of 550 significant digits ($N_{s} = 550$) mentioned above, in the accuracy of at least 8 significant digits in the whole interval of time $t\in[0,10000]$. Especially, it takes only 250 seconds (i.e. about 4 minutes), say, only 4.3\% of the CPU time (i.e. 5804 seconds) of the previous CNS algorithm with a {\em fixed} time step and a {\em fixed} value of $N_{s}$ for multiple-precision. This further verifies the high computational efficiency of the self-adaptive CNS algorithm mentioned in \S~2.

\subsection{Self-adaptive CNS for three-body problem}

Here let us consider the well-known three-body problem \cite{poincare1890probleme,diacu1999celestial,henon1964applicability,valtonen2006three}, i.e. the motion of three celestial objects/bodies under their mutual gravitation. Let $x_1$, $x_2$, $x_3$ denote three Cartesian coordinates and $\mathbf{r}_i=(x_{1,i},x_{2,i},x_{2,i})$ denotes the corresponding position vector of the $i$th body. Considering Newton's Law of Gravitation, the motion of three bodies is governed by the following non-dimensional equation
\begin{equation}
\ddot{x}_{k,i}=\sum_{j=1,j\neq i}^{3}\rho_j\frac{x_{k,j}-x_{k,i}}{R^3_{i,j}}, \hspace{1.0cm} k=1,2,3,    \label{3b}
\end{equation}
where
\begin{equation}
R_{i,j}=\left[ \sum_{k=1}^{3}(x_{k,j}-x_{k,i})^2 \right]^{\frac{1}{2}}
\end{equation}
and
\begin{equation}
\rho_i=\frac{m_i}{m_1}, \hspace{1.0cm} i=1,2,3
\end{equation}
denotes the ratio of mass, in which $m_i$ denotes the mass of the $i$th body.

Similarly, in the frame of the CNS, the background numerical noise (i.e. truncation error and round-off error) of solving the three-body problem (\ref{3b}) can be decreased under a {\em required} tiny level by means of choosing a large {\em enough} order $M$ of the Taylor expansion and a large {\em enough} number $N_s$ of significant digits for multiple-precision. For more details, please refer to Liao \cite{liao2014physical}.

Without loss of generality, in this paper we follow Liao \cite{liao2014physical} to consider the motion of three bodies with the initial positions
\begin{equation}
\mathbf{r}_1=(0,0,-1)+d\mathbf{r}_1, \hspace{1.0cm} \mathbf{r}_2=(0,0,0), \hspace{1.0cm} \mathbf{r}_3=-(\mathbf{r}_1+\mathbf{r}_2),    \label{3b_ini}
\end{equation}
as well as the initial velocities
\begin{equation}
\dot{\mathbf{r}}_1=(0,-1,0), \hspace{1.0cm} \dot{\mathbf{r}}_2=(1,1,0), \hspace{1.0cm} \dot{\mathbf{r}}_3=-(\dot{\mathbf{r}}_1+\dot{\mathbf{r}}_2),    \label{3b_v}
\end{equation}
where $d\mathbf{r}_1=\delta\,(1,0,0)$ denotes the micro-level physical uncertainty with $\delta=10^{-60}$. For simplicity, we consider the case of equal masses, say, $\rho_j=1$ with $j=1,2,3$.

It is easy for us to know that the maximum Lyapunov exponent of the above-mentioned three-body problem is about 0.168, which gives us the corresponding noise-growing exponent $\kappa \approx 0.168$ in (\ref{noise_exp}). In this case, if $\varepsilon_{0} = 10^{-N_{s}}$, we have the numerical noise evolution
\[ \varepsilon(t) \approx \varepsilon_{0} \exp(0.168\hspace{0.3mm}t) \approx 10^{-(N_{s}-0.073\hspace{0.3mm}t)}. \]
If our CNS simulation should be in the accuracy of at least 11 significant digits in the whole interval of $t\in[0,1000]$, we have $T_{c} = 1000$ and
\[  -(N_{s} - 0.073 \hspace{0.3mm} T_{c}) \leq -11,  \]
which gives $N_{s} \geq 84$, indicating that we should choose $N_{s} = 84$.

First, following Liao \cite{liao2014physical} who used a constant time step, we obtain a reproducible/convergent simulation of the three-body problem (\ref{3b})-(\ref{3b_v}) in $t\in[0,1000]$ by means of a CNS algorithm using the 45th-order Taylor expansion ($M=45$) with a fixed time step $\Delta t = 0.01$ in the multiple precision of 84 significant digits ($N_{s} = 84$). In fact, this convergent simulation result agrees in the accuracy of at least 11 significant digits in the whole interval of time $t\in[0,1000]$ compared with the benchmark solution (with the even smaller background numerical noise) given by another CNS using the 60th-order Taylor expansion ($M=60$) with a fixed time step $\Delta t = 0.01$ in the multiple precision of 100 significant digits ($N_{s} = 100$). It takes 1327 seconds (i.e. about 22 minutes) using Intel's CPU (Xeon Silver 4114) on our local cluster.

Then, we apply the strategy of keeping a balance between truncation error and round-off error (mentioned in \S~2.1) to increase the computational efficiency of the CNS algorithm. The variable stepsize (VS) scheme is applied with an optimal time step determined by (\ref{tol}), where $x_i$ is replaced by $x_{k,i}$ for this three-body problem. Considering that (\ref{lorenz_tol}) is used to keep the balance between the round-off error and the truncation error, and there is no parallel technology applied in the CNS algorithm for solving the three-body problem (\ref{3b})-(\ref{3b_v}), here we adopt the optimal order of Taylor expansion \cite{jorba2005software}
\begin{equation}
M=\left\lceil 1.15\hspace{0.3mm}N_s+1 \right\rceil.    \label{3b_M}
\end{equation}
In this way, we can control the background numerical noise $\varepsilon_{0}$ by choosing the number $N_s$ of significant digits for multiple-precision, say, $\varepsilon_{0}$ is at the level of $10^{-N_{s}}$.

By means of the CNS algorithm described in \S~2.1 using a {\em fixed} value of $N_{s} = 84$ for multiple-precision, $tol = 10^{-84}$ for the allowed tolerance, $M=\left\lceil 1.15\hspace{0.3mm}N_s+1 \right\rceil=98$ for the order of Taylor expansion, and an optimal time step given by (\ref{tol}), we obtain a reproducible/convergent numerical simulation of the three-body problem (\ref{3b})-(\ref{3b_v}) in $t\in[0,1000]$, and it takes 370 seconds (i.e. about 6 minutes) using Intel's CPU (Xeon Silver 4114) on our local cluster, say, only 28\% CPU time of the previous CNS algorithm (i.e. 1327 seconds) with a {\em fixed} time step. Furthermore, this CNS result is in the accuracy of more than 11 significant digits in the whole interval of time $t\in[0,1000]$, compared with the above-mentioned CNS benchmark solution.

\begin{figure}[h!]
    \begin{center}
        \begin{tabular}{cc}
            \includegraphics[width=3.2in]{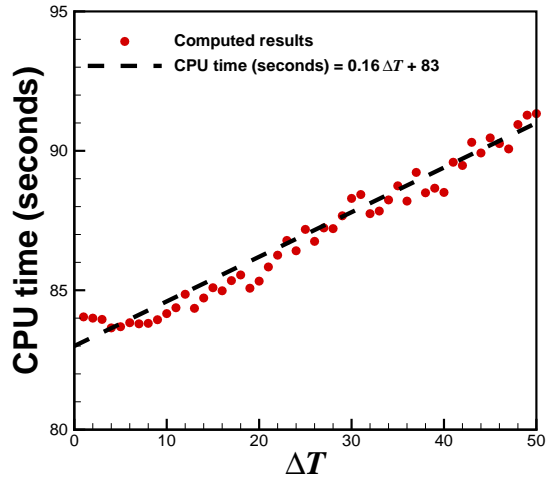}
        \end{tabular}
    \caption{CPU times of the self-adaptive CNS algorithm for the three-body problem (\ref{3b})-(\ref{3b_v}) in $t\in[0,1000]$, given by different values of $\Delta T$, using $tol = 10^{-N_{s}}$ and $M=\left\lceil 1.15\hspace{0.3mm}N_s+1 \right\rceil$ with the optimal time step (\ref{tol}), where $N_{s}$ is determined by (\ref{SA-Ns-prac-3b}) with taking $T_c=1000$. Red circle: computed results; Black dashed line: CPU time (seconds) equals to $0.16\hspace{0.3mm}\Delta T+83$.}    \label{CPU_tao_3b}
    \end{center}
\end{figure}

In addition, the computational efficiency can be further increased by means of the self-adaptive CNS algorithm described in \S~2.2. Substituting $\kappa=0.168$ into (\ref{SA-Ns}) and choosing $\gamma=1.1$, $\varepsilon_c=10^{-2}$, we have the following relationship
\begin{equation}
N_s = \left\lceil 0.08\hspace{0.3mm}(T_c-t^*)+2 \right\rceil,    \label{SA-Ns-prac-3b}
\end{equation}
where $t^*=n\,\Delta T$ with the non-negative integer $n$. In practice, there is $T_c-t^*>200$ for the high enough remaining precision, say, the value of $N_s$ is stopped decreasing when $t>800$ for the long time simulation with $t\in[0,1000]$ in this section. Using different values of $\Delta T$, the corresponding CPU times of this self-adaptive CNS algorithm with the adjustable multiple-precision (MP) and an optimal variable time step, for the three-body problem (\ref{3b})-(\ref{3b_v}) in $t\in[0,1000]$, i.e. $T_c=1000$, are illustrated in Fig.~\ref{CPU_tao_3b}.
It indicates that there is an approximate linear relationship that the required CPU time (seconds) equals to $0.16\hspace{0.3mm}\Delta T+83$. Since the slope 0.16 of this linear relationship is rather small, it once again confirms the conclusion that the required CPU time of the above-mentioned self-adaptive CNS algorithm is not very sensitive to the value of $\Delta T$. Thus, in practice we can choose $\Delta T=0.5\%\hspace{0.3mm}T_c=5$ for the relatively higher computational efficiency.

For the three-body problem (\ref{3b})-(\ref{3b_v}) in $t\in[0,1000]$, using $tol = 10^{-N_{s}}$ and $M=\left\lceil 1.15\hspace{0.3mm}N_s+1 \right\rceil$ (according to (\ref{lorenz_tol}) and (\ref{3b_M}), respectively) with the optimal time step (\ref{tol}), where $N_{s}$ is determined by (\ref{SA-Ns-prac-3b}) with taking $T_c=1000$ and $\Delta T=5$, we successfully obtain the {\em same} reproducible/convergent numerical simulation by means of the above-mentioned self-adaptive CNS algorithm using Intel's CPU (Xeon Silver 4114) on our local cluster, which agrees with the benchmark solution given by another CNS (with the even smaller background numerical noise), using the 60th-order Taylor expansion ($M=60$) with a fixed time step $\Delta t = 0.01$ in the multiple precision of 100 significant digits ($N_{s} = 100$) mentioned above, in the accuracy of at least 11 significant digits in the whole interval of time $t\in[0,1000]$. Especially, it takes only 84 seconds (i.e. less than 2 minutes), say, only 6.3\% of the CPU time (i.e. 1327 seconds) of the previous CNS algorithm with a {\em fixed} time step and a {\em fixed} value of $N_{s}$ for multiple-precision. This result also verifies the high computational efficiency of the self-adaptive CNS algorithm mentioned in \S~2.

\subsection{Self-adaptive CNS for spatiotemporal chaos}

Let us consider here a spatiotemporal chaos, i.e. a chain of pendulums coupled through the elastic restoring force, governed by the damped driven sine-Gordon equation \cite{chacon2008spatiotemporal, keller2013surveys, ferre2017localized}:
\begin{equation}
u_{tt}(x,t)=u_{xx}(x,t)-sin[u(x,t)]-\alpha\hspace{0.2mm}u_t(x,t)+\Gamma\hspace{0.2mm}sin(\omega\hspace{0.2mm}t-\lambda\hspace{0.2mm}x),    \label{SGE}
\end{equation}
subject to a periodic boundary condition
\begin{equation}
u(x+l,t)=u(x,t),    \label{SG_periodic}
\end{equation}
where the subscript represents the spatial/temporal partial derivative, $x$ and $t$ denote the variables in the spatial and temporal dimensions, $u(x,t)$ represents the angle of a pendulum, $\alpha$ denotes a constant related to the damped friction, $\Gamma$ denotes a constant related to the external force field, $\omega$ is the temporal frequency and $\lambda=2\pi/l$ is the spatial frequency, $l$ denotes the total calculating length of the system, respectively. Without loss of generality, we follow Chac\'{o}n et al. \cite{chacon2008spatiotemporal} to consider the following case
\begin{equation}
\omega=\frac{3}{5}, \;\; \alpha=\frac{1}{10}, \;\;  \Gamma=\frac{461}{500}, \;\;  l=500, \;\;  \lambda=\frac{2\pi}{l}=\frac{\pi}{250},    \label{SG_parameters}
\end{equation}
with the initial condition
\begin{equation}
u(x,0)=0, \;\;  u_t(x,0)=0.    \label{SG_initial}
\end{equation}
As reported by Qin \& Liao \cite{qin2020influence}, the above-mentioned model corresponds to a spatiotemporal chaos, whose statistics are extremely sensitive to a small disturbance: such kind of chaos belongs to the so-called ultra-chaos, which is in a higher level of disorders than a normal-chaos, as reported by Liao \& Qin \cite{AAMM-14-799}.

Similarly, the CNS algorithm for the sine-Gordon equation (\ref{SGE})-(\ref{SG_initial}) is also based on a high {\em enough} order of Taylor expansion in the temporal dimension for decreasing the temporal truncation error under a {\em required} tiny level, but combined with a high {\em enough} order of the spatial Fourier expansion for a fine {\em enough} spatial discretization for decreasing the spatial truncation error under a {\em required} tiny level. First, the spatial interval $x\in[0,l\hspace{0.2mm})$ is discretized uniformly by $N$ equidistant points, say, $x_k=l\,k/N$, where $k=0,1,2,...,N-1$ and $x_k$ denotes the $k$th discrete points in the physical space. Note that the parallel technology is applied in the spatial discretization. Then, in the temporal dimension, the $M$th-order Taylor expansion method is used in the temporal interval $[t, t+\Delta t]$, say,
\begin{equation}
u(x_k,t+\Delta t)\approx u(x_k,t)+\sum_{m=1}^{M}u^{[m]}(x_k,t)\,(\Delta t)^m, \hspace{1.0cm} 0\leq  k \leq N,     \label{Taylor_SG}
\end{equation}
where $\Delta t$ is the time step and
\begin{equation}
u^{[m]}(x_k,t)=\frac{1}{m!}\frac{\partial^m u(x_k,t)}{\partial t^m}.
\end{equation}
From (\ref{Taylor_SG}), we have the first-order derivative with respect to time:
\begin{equation}
u_t(x_k,t+\Delta t) = u^{[1]}(x_k,t+\Delta t)\approx u^{[1]}(x_k,t)+\sum_{m=1}^{M}\,(m+1)\,u^{[m+1]}(x_k,t)\,\left(\Delta t\right)^m.    \label{Taylor1_SG}
\end{equation}
The high-order temporal derivatives in (\ref{Taylor_SG}) and (\ref{Taylor1_SG}) can be obtained via differentiating both sides of Eq.~(\ref{SGE}) with respect to $t$, while the corresponding spatial derivatives are approximated by means of the $N$th-order Fourier spectral expression, i.e.
\begin{equation}
u^{[m]}(x,t)\approx \frac{1}{2}\,a_{m,0}(t)+\sum_{n=1}^{\frac{N}{2}-1}\left[a_{m,n}(t)\,cos(\lambda\hspace{0.2mm}nx)+b_{m,n}(t)\,sin(\lambda\hspace{0.2mm}nx)\right]
+a_{m,\frac{N}{2}}(t)\,cos\hspace{0.1mm}\left(\frac{\lambda\hspace{0.2mm}N\hspace{-0.1mm}x}{2}\right),    \label{Fourier}
\end{equation}
where the Fast Fourier Transform (FFT) algorithm as well as parallel technology can be adopted. For more details, please refer to Qin \& Liao \cite{qin2020influence}. Besides, all physical and numerical variables/parameters are in the multiple precision with a large {\em enough} number $N_s$ of significant digits so as to decrease the round-off error under a {\em required} tiny level. As reported by Hu \& Liao \cite{hu2020risks} and Qin \& Liao \cite{qin2020influence}, this kind of parallel CNS algorithm in physical space for spatio-temporal chaos has much higher computational efficiency than the previous CNS algorithm in spectrum space \cite{lin2017origin}. To further increase the computational efficiency, Qin \& Liao \cite{qin2020influence} applied the VS scheme in the temporal dimension with a given allowed tolerance $tol$ for solving the governing equation, using an optimal time step determined by:
\begin{equation}
\Delta t=min\hspace{0.2mm}\Bigg(\frac{tol^{\frac{1}{M}}}{\|u^{[M-1]}(x_k,t)\|_\infty^{\frac{1}{M-1}}},
\frac{tol^{\frac{1}{M+1}}}{\|u^{[M]}(x_k,t)\|_\infty^{\frac{1}{M}}}\Bigg),    \label{SG_tol}
\end{equation}
where $\|~\|_\infty$ is the infinite norm for the variable $x_k$. We adopted the empirical formula $M=\left\lceil-\log_{10}(tol)-10\right\rceil$ to determine a proper order of Taylor expansion for the high calculating efficiency \cite{qin2020influence}. In addition, we use (\ref{lorenz_tol}) to balance the round-off error at the same level of the temporal truncation error. In this way, one can control the background numerical noise $\varepsilon_{0}$ only by means of choosing the number $N_s$ of significant digits for multiple precision.

In this way, Liao \& Qin \cite{AAMM-14-799} obtained a convergent chaotic simulation $u(x,t)$ of the damped driven sine-Gordon equation (\ref{SGE})-(\ref{SG_initial}) in $t\in[0,3600]$ by means of  a parallel algorithm of the CNS using $N = 2^{16}=65536$ and $N_s = 230$, corresponding to $tol=10^{-230}$ and $M=220$ according to (\ref{lorenz_tol}) and (\ref{SG_M}), respectively, with the optimal time step via (\ref{SG_tol}). It took 202.6 hours (about 8 days and 11 hours) using 256 Intel's CPUs (Xeon Silver 4114) on our local cluster.
To confirm its convergence in the whole interval $t\in[0,3600]$, say, $T_c=3600$, Liao \& Qin \cite{AAMM-14-799} obtained another CNS result $u'(x,t)$ with the even smaller background numerical noise using the same $N = 65536$ but $N_s = 240$, corresponding to $tol=10^{-240}$ and $M=230$ according to (\ref{lorenz_tol}) and (\ref{SG_M}), respectively, with the optimal time step via (\ref{SG_tol}). The deviation of $u(x,t)$ from $u'(x,t)$ is given by
\begin{equation}
\varepsilon(t) = \frac{\sum\limits_{n=0}^{\frac{N}{2}}\left|(c_n')^2-(c_n)^2\right|}{\sum\limits_{n=0}^{\frac{N}{2}}|c'_n|^2},    \label{delta_s}
\end{equation}
where $c_{n}$ and $c'_{n}$ are the the complex coefficients of the spatial Fourier expansion of $u(x,t)$ and $u'(x,t)$ at a given time $t$, respectively. It was found that the deviation evolves in a power law
\begin{equation}
\varepsilon(t) \approx \varepsilon_{0 }\exp\hspace{0.2mm}(\kappa\,t),    \label{SGE_noise_exp}
\end{equation}
where $\varepsilon_{0} = 10^{-N_{s}}$ is the background numerical noise and $\kappa \approx 0.14$ is the noise-growing exponent. For more details, please refer to Liao \& Qin \cite{AAMM-14-799}.

Similarly, to further increase the computational efficiency, here we adopt the self-adaptive CNS algorithm described in \S~2.2. Substituting $\kappa=0.14$ into (\ref{SA-Ns}) and choosing $\varepsilon_c=10^{-2}$, we have the following relationship
\begin{equation}
N_s = \left\lceil \frac{\gamma\,(T_c-t^*)}{16.4}+2 \right\rceil,    \label{SA-Ns-prac-EX2}
\end{equation}
where $t^*=n\,\Delta T$ with the non-negative integer $n$. In practice, $\Delta T=0.5\%\hspace{0.3mm}T_c=18$ is chosen for the higher computational efficiency and there is $T_c-t^*>600$ for the high enough remaining precision, say, the value of $N_s$ is stopped decreasing when $t>3000$ for the long time simulation with $t\in[0,3600]$ in this section. Considering that (\ref{lorenz_tol}) is used to keep the balance between the round-off error and the truncation error, and there is no parallel technology applied in the CNS algorithm for solving the sine-Gordon equation (\ref{SGE})-(\ref{SG_initial}), here we also adopt the optimal order of Taylor expansion \cite{jorba2005software}
\begin{equation}
M=\left\lceil 1.15\hspace{0.3mm}N_s+1 \right\rceil.    \label{SG_M}
\end{equation}

We adopt the self-adaptive CNS algorithm mentioned above for solving the sine-Gordon equation (\ref{SGE})-(\ref{SG_initial}), using $N = 65536$ for the spatial discretization, the allowed tolerance $tol = 10^{-N_s}$ of governing equations via (\ref{lorenz_tol}), and the order $M=\left\lceil 1.15\hspace{0.3mm}N_s+1 \right\rceil$ of Taylor expansion via (\ref{SG_M}) with the optimal time step (\ref{SG_tol}), respectively, where the self-adaptive number of $N_{s}$ for multiple-precision is determined by (\ref{SA-Ns-prac-EX2}) with taking $\gamma=1.2$, $T_c=3600$ and $\Delta T=18$.
Here it should be emphasized that, according to the definition (\ref{delta_s}), the power law (\ref{SGE_noise_exp}) means an averaged evolution of deviation, and the real deviations at different discrete points in the physical space have fluctuations compared with this averaged deviation. Thus we choose a relatively large safety factor $\gamma=1.2$ for enough precision.

Note that the {\em same} reproducible/convergent numerical result in the averaged accuracy of 5 significant digits in the whole interval of time $t\in[0,3600]$ (compared with the benchmark solution given by another CNS using the same $N = 65536$ but a fixed value of $N_s = 240$, corresponding to $tol=10^{-240}$ and $M=230$ according to (\ref{lorenz_tol}) and (\ref{SG_M}), respectively) is obtained, by means of the self-adaptive CNS algorithm mentioned above, which takes 76.9 hours (i.e. about 3 days and 5 hours) using 256 Intel's CPUs (Xeon Silver 4114) on our local cluster, say, only about 38\% of the CPU time (i.e. 202.6 hours) of the previous CNS algorithm. This illustrates that the self-adaptive CNS algorithm described in \S~2 can indeed greatly increase the computational efficiency for a spatiotemporal chaos.

\section{Conclusion}

The background numerical noise $\varepsilon_{0} $ is determined by the maximum of truncation error and round-off error. For a chaotic dynamical system, the numerical error $\varepsilon(t)$ grows exponentially, say, $\varepsilon(t) = \varepsilon_{0}  \exp(\kappa\,t)$, where $\kappa>0$ is the so-called noise-growing exponent. This is the reason why one can not gain a convergent simulation of chaotic systems in a long enough interval of time by means of traditional algorithms in double precision, since the background numerical noise $\varepsilon_{0}$ might stop decreasing because of the use of double precision.
This restriction can be overcome by means of the clean numerical simulation (CNS) \cite{Liao2009, Liao2013, Liao2014, LIAO2014On}, which can decrease the background numerical noise $\varepsilon_{0}$ to any a required tiny level. A lot of successful applications show the novelty and validity of the CNS \cite{hu2020risks, qin2020influence, xu2021accurate, AAMM-14-799, liao2015inherent, lin2017origin, Li2017More, li2018over, li2019collisionless, li2021one, qin_liao_2022}. In this paper, we propose some strategies to greatly increase the computational efficiency of the CNS algorithms for chaotic dynamical systems. It is highly suggested to keep a balance between truncation error and round-off error and besides to progressively enlarge the background numerical noise $\varepsilon_{0}$, since the exponentially increasing numerical noise $\varepsilon(t)$ is much larger than it. To illustrate the validity of our strategies, we apply the CNS algorithm combined with the self-adaptive precision to some chaotic dynamical systems, such as the Lorenz system, the hyper-chaotic R\"{o}ssler system, the three-body problem, and a spatiotemporal chaos governed by the damped driven sine-Gordon equation. All of our results indicate that the self-adaptive CNS algorithm can indeed greatly increase the computational efficiency for chaotic systems.

\section*{Declaration of competing interest}
The authors declare that they have no known competing financial interests or personal relationships that could have appeared to influence the work reported in this paper.

\section*{Acknowledgments}
This work is partly supported by National Natural Science Foundation of China (No. 12272230) and Shanghai Pilot Program for Basic Research - Shanghai Jiao Tong University (No. 21TQ1400202). The parallel algorithms for the Lorenz system in this paper were performed on TH-2 at National Supercomputer Centre in Guangzhou, China.

\section*{References}
\bibliography{cns}

\end{document}